\documentstyle[prl,aps,floats]{revtex} 

\begin{document}
\draft

\twocolumn[\hsize\textwidth\columnwidth\hsize\csname @twocolumnfalse\endcsname

\title{Cold Collision Frequency Shift of an Optical Spectrum of a Trapped Gas}
\author{ 
Mehmet~\"O.~Oktel, Thomas~C.~Killian$^\ast$, Daniel~Kleppner$^\ast$, 
L.~S.~Levitov
}

\address{Department of Physics,
Massachusetts Institute of Technology,
Cambridge, MA 02139\\
(*) Also, Research Laboratory of Electronics, MIT
 }
\maketitle

\begin{abstract}
We develop an exact sum rule that relates the spectral shift of a trapped
gas undergoing cold collisions to measurable quantities of the system.
The method demonstrates the dependence of the cold collision frequency
shift on the quantum degeneracy of the gas and facilitates extracting
scattering lengths from the data. We apply the method to analyzing
spectral data for magnetically trapped hydrogen atoms and determine
the value of the $1S-2S$ scattering length.
   \end{abstract}
\pacs{PACS: 03.75.Fi, 32.80.Pj }
 ]

The broadening and shifting of spectral lines of a gas by collisions was
among the earliest discoveries in the development of high precision
spectroscopy\cite{Michelson}.  
The  pressure shift, which originates in
interatomic
perturbations\cite{tsc75}, is particularly simple to interpret at low
temperatures where
the thermal de  Broglie wavelength
$\Lambda_{T}=h/(2\pi m k_B T)^{1/2}$ is much larger than the scattering length
$a$\cite{wbz99} and
the interactions arise only through $s$-wave scattering. In this {\it cold
collision} regime, the frequency shift is much larger than the level
broadening.

The theory of the cold collision shift has been developed to interpret
hyperfine transitions in cryogenic hydrogen masers and
laser cooled atomic fountains\cite{kvg97}. In this work we study
the shift for optical excitation in a system that can
be quantum degenerate, and apply the results to data on $1S-2S$ two--photon
excitation of trapped atomic hydrogen\cite{kfw98}.

For the case of a homogeneous sample of density $n$, 
and a coherent, weak excitation
that couples two inner states of the atoms, we find
  \begin{equation}
\hbar\Delta\omega_{\rm coll} = g_2 (\lambda_{12} -\lambda_{11})n
\ ,\qquad \lambda_{\alpha\beta}=
4\pi \hbar^2 a_{\alpha\beta}/ m \ .
  \label{boltsshift}
  \end{equation}
Here $g_2$ is the equal point value 
of the second order correlation function \cite{gfactor}, 
$g_2 \equiv g^{(2)}(\vec{r}=0)$, the state 1(2) is
the ground(excited) state,  and $a_{\alpha\beta}$ is the $s$-wave scattering 
length for ${\alpha-\beta}$ collisions. 

Equation (\ref{boltsshift}) shows that quantum correlations  
in the system are manifest in the collision shift.
For a uniform Bose gas in thermal equilibrium $g_2= 2 - (n_{\rm BEC}/n)^2$ 
\cite{LL},
where $n_{\rm BEC}$ is the density of condensed 
atoms. Above the condensation temperature, when $n_{\rm BEC} = 0$, $g_2$ 
equals 2,
in which case Eq.(\ref{boltsshift}) is in agreement with previous
work \cite{kvg97}. At zero temperature, for a pure condensate with
$n_{\rm BEC}=n$, the collision shift is half of the shift for 
a non-condensed gas. Equation (\ref{boltsshift}) generalizes 
the result \cite{kvg97} 
to $T<T_{\rm BEC}$ and relates the spectral shift to the condensate fraction.

It is quite remarkable that the factor $g_2$ in Eq.(\ref{boltsshift})
multiplies {\it both} $\lambda_{12}$ and $\lambda_{11}$. 
This results from correlations between 
an excited atom and other atoms. 
During the excitation, the internal states of the atoms
are rotated: $\cos\theta(t)|1S\rangle +e^{-i\phi(t)}\sin\theta(t)|2S\rangle$.
The angles $\theta(t)$, $\phi(t)$ depend on laser power and on 
the atom's trajectory in the laser field, specific for each atom. 
However, for small excitation power, the angle $\theta(t)$ is small, 
and thus the internal 
states of all atoms remain nearly identical while the laser is on,
even if the excitation field is spatially nonuniform. 
Therefore, during the excitation the atoms interact as identical particles.
This causes the short range statistical correlations in the 
initial state to be replicated in the excited state of the gas,
which results in the statistical factor $g_2$ in the 
first term of Eq.(\ref{boltsshift}). 

The transfer of spatial correlations to the excited state is not limited to 
weak excitation. For the case of strong excitation, spatial correlations 
in the ground state will also be transferred to the excited state, 
but only provided the excitation scheme is coherent. The difference
between coherent and incoherent cases can be seen from comparing two examples, 
the coherent superposition of the ground and excited states obtained, 
e.g., by a $\pi/2$ pulse, and the incoherent mixture state resulting 
from saturating the Rabi transition. 
These states will both have equal populations in the two internal
states, but quite different correlations.
In the former case of a pure internal state
the spatial correlation will be the same as for the ground state of
indistinguishable particles. In the latter case of a mixed state 
the correlations will be reduced. Consequently, the correlation energy
of the first state will exceed that of the second state by the 
factor $g_2$. 

To emphasize the non-trivial character of the result Eq.(\ref{boltsshift}), 
let us point out that $\hbar\Delta\omega_{\rm coll}$ differs from 
the thermodynamic work needed to transfer one atom from the state 
1 to the state 2. The latter work, calculated by removing one atom 
from the sample, and then introducing an atom in state 2 from far away, 
ignoring entropy, 
is given by $(\lambda_{12} - g_2 \lambda_{11})n$. Here $\lambda_{12} n$ is 
the energy of interaction of the excited atom with the atoms in the state 1, 
and  $g_2 \lambda_{11} n$ is
the chemical potential of a Bose gas.
The key difference between this process 
and optical excitation, resulting in the different dependence on $g_2$, 
is the incoherence of the state of the added atom
with the initial state of the sample.

To calculate the full optical spectrum shape of a trapped gas in the cold collision
regime, other factors
would have to be considered in addition to the effects
of statistical correlations. Optical coherence
can be lost via dephasing elastic collisions, 
giving rise to collisional broadening. 
One would have also to take into account atomic motion in the trap and the
effects of the inhomogeneous density distribution in the sample, 
especially in the Bose-Einstein condensate. 
In addition, the interaction may give rise to a doublet structure of the
spectrum\cite{ole99}. Altogether, these effects can lead to a complicated 
broadened spectrum with asymmetric lines\cite{kfw98,fkw98,tcknew}. 
However, we demonstrate below that the spectrum's center of mass 
obeys a simple and exact sum rule and is insensitive 
to these additional effects. 

We lay out the theory of the shift by deriving a sum rule (Eq.(\ref{sum-total}))
that relates the center of mass of the observed spectrum
to measurable experimental parameters. The sum rule 
bridges between the uniform density result Eq.(\ref{boltsshift}) 
and experimentally measured spectra. The sum rule accounts for 
all interactions between atoms occuring in the $s-$wave scattering channel,
which includes the $s-$wave collisional broadening. It follows from the sum rule
that collisional broadening as well as the time of flight broadening
resulting from atomic motion in the trap do not contribute to the spectral 
shift. At the same time, the effects on the shift of 
inhomogeneity in the gas density and non-uniformity in the excitation field
are expressed in the sum rule Eq.(\ref{sum-total}) in an exact 
and straightforward way. 

We start by considering a homogeneous Bose gas and derive Eq.(\ref{boltsshift}). 
Then for the realistic situation of a trapped  gas sample 
we derive the sum rule Eq.(\ref{sum-total}), 
a generalization of Eq.(\ref{boltsshift}). The sum rule is exact and general,
applicable both to Doppler-free and Doppler-sensitive spectra. 
Finally we apply the sum rule to experimental data
on the spectrum of cold trapped hydrogen to calculate 
the $1S-2S$ scattering length for hydrogen. 

\noindent{\bf The system:}
To provide the context for the theory, we briefly describe the experimental
situation.  The temperature of the
hydrogen is $30-100\,\mu{\rm K}$, well below 
the cold collision  threshold $T\simeq 1~K$\cite{wbz99}. 
The atoms are spin polarized and interact in the triplet channel.
Calculated values of 
the $1S-1S$ and $1S-2S$ triplet scattering lengths are 
$a_{11}=0.0648$~nm\cite{jdk95} and $a_{12} = -2.3$~nm\cite{jdd96}. 
We neglect $2S-2S$ scattering
because the excitation rate is assumed low (in the experiment 
typically $10^{-4}$ of the atoms are excited) 
so the background gas is essentially
pure $1S$. Since $|a_{12}|\gg a_{11}$,
collisions between $1S$ and $2S$ atoms dominate the shift, which is to the red.

Each atom will be in some superposition of 
the ground state $1S$ and the excited state $2S$. 
In the second quantization formalism, the atoms are 
described by the canonical Bose
operators $\psi_1(r)$ and $\psi_2(r)$.
The Hamiltonian is
${\cal H}={\cal H}_0+{\cal H}_{\rm int}$, where ${\cal H}_0$ describes 
atoms freely moving in the trap, and ${\cal H}_{\rm int}$ is the interaction 
term:
   \begin{eqnarray}\label{Ham}
{\cal H}_0 &=&\int\sum\limits_{\alpha=1,2} \psi^+_\alpha(r)\left(
-\frac{\hbar^2\nabla^2}{2m} + U(r)\right)\psi_\alpha(r)
\ d^3r\ ,
  \\
\label{Ham_int}
{\cal H}_{\rm int} &=& \frac{1}{2} \int\sum\limits_{\alpha,\beta=1,2}
\lambda_{\alpha\beta}\psi^+_\alpha(r)\psi^+_\beta(r)\psi_\beta(r)\psi_\alpha(r)
\ d^3r\ .
   \end{eqnarray}
Here $U(r)$ is the trap potential 
(essentially the same for the $1S$ and the $2S$ states).

Inelastic collisions, such as collisions in which the hyperfine level of
one or both of the colliding partners changes, may contribute additional
shifts which are not accounted for in this formalism. However, these effects, 
as well
as the three-body collision effects, are small in the experiment and can be 
neglected.

The two-photon $1S-2S$ spectrum consists of Doppler-free and 
Doppler-sensitive excitations. In the Doppler-free situation, 
the transition results from absorbing two counter-propagating photons 
with equal frequencies and zero net momentum. In the absence of interactions, 
the resonance condition is
$2\omega_{\rm laser}=\omega_0$, where $\omega_0$ corresponds 
to the resonance of a single free atom. In the Doppler-sensitive situation, 
the transition is caused by two photons propagating in the same direction. 
For a free atom, the resonance frequency is shifted by the recoil energy:
$2\hbar\omega_{\rm laser}=\hbar\omega_0+(2k)^2/2m$, where $k=\hbar\omega/c$ 
is photon momentum and $m$ is the atom mass.

Radiative excitation in a many 
particle system
is described by adding to the Hamiltonian (\ref{Ham}),(\ref{Ham_int}) the term 
  \begin{equation}\label{H_ext}
  {\cal H}_{\rm rad} =
\int d^3r\left( A(r) e^{-i\omega t} \psi^+_2(r)\psi_1(r)\ +\ {\rm 
h.c.}\right)\ ,
  \end{equation}
where  $\omega=2\omega_{\rm laser}-\omega_0$.
The two--photon excitation field $A(r)$ is equal, up to a constant factor, 
to the square of the electric field.  
Spatial variation of $A(r)$ in the Doppler-free case occurs on a scale set by  
the focused laser beam diameter, 
and in the Doppler-sensitive case is given by $\tilde A(r)\cos(2kr+\phi(r))$, where 
$\tilde A(r)$ and $\phi(r)$ are slowly varying functions.

\noindent{\bf A tutorial example:} Before discussing the general case, 
here we derive the mean frequency shift for the Doppler-free transition caused by 
a uniform excitation field $A(r)=A_0$, ignoring the $1S-1S$ 
interactions ($\lambda_{11}=0$). To that end, consider a gas of $N$ 
atoms confined in a box of volume $V$. Since we ignore the $1S-1S$ interaction,
the many body state ground state of the system $\Phi_0$
is simply a symmetrized product of single particle states. It can be
characterized by occupation numbers $n_j$ of 
the single particle plane wave states 
$V^{-1/2}e^{ik_jr}$, $\sum_j n_j=N$. Initially, the 
internal state of all atoms is $1S$. 

The excited state, to lowest order in the excitation, 
is given by $\Phi_1={\cal H}_{\rm rad}\Phi_0$. 
We consider the norm 
$\parallel\Phi_1\parallel^2$ and the expectation value of the interaction
$\langle\Phi_1|{\cal H}_{\rm int}|\Phi_1\rangle$. The ratio of these quantities
gives the mean frequency shift. Because $\Phi_0$ is
the product of plane wave states in a box, the frequency shift 
can be evaluated exactly.  

The norm $\langle \Phi_1 |\Phi_1\rangle$ of the excited state is given by
  \begin{equation}\label{norm1}
\parallel\Phi_1\parallel^2=
|A_0|^2\int
\langle\Phi_0|
\psi^+_1(r)\psi_2(r)\psi^+_2(r')\psi_1(r')
|\Phi_0\rangle d^3r d^3r'
\ .
  \end{equation}
To evaluate the norm one first puts the operators 
$\psi_2(r)$ and $\psi^+_2(r')$
 in (\ref{norm1}) in normal order by using the commutation relation 
$[\psi_2(r),\psi^+_2(r')]=\delta(r-r')$. 
Noting that $\psi_2(r)|\Phi_0\rangle=0$, the norm 
is given by
  \begin{equation}\label{norm}
\parallel\Phi_1\parallel^2 = |A_0|^2\int
\langle\Phi_0|
\psi^+_1(r)\psi_1(r)
|\Phi_0\rangle d^3r =|A_0|^2N
\ .
  \end{equation}
To obtain the frequency shift $\Delta\omega_{\rm coll}$, 
we consider the expectation value
$\langle\Phi_1|{\cal H}_{\rm int}|\Phi_1\rangle$,
keeping in ${\cal H}_{\rm int}$ only the $1S-2S$ interaction $\lambda_{12}$.
After arranging in  normal order, as in  the calculation 
of the norm $\parallel\Phi_1\parallel^2$, one has
  \begin{equation}\label{H-mean}
\langle\Phi_1|{\cal H}_{\rm int}|\Phi_1\rangle =
\lambda_{12}|A_0|^2\int
\langle\Phi_0|
\psi^+_1(r)\psi^+_1(r)\psi_1(r)\psi_1(r)
|\Phi_0\rangle d^3r 
\ .
  \end{equation}
Evaluating the expectation value for $\Phi_0$ 
chosen as a product of plane wave states, 
one expresses Eq.(\ref{H-mean}) in terms of 
the occupation numbers of the ground and excited states as
  \begin{equation}\label{H-mean-result}
\langle\Phi_1|{\cal H}_{\rm int}|\Phi_1\rangle =
\lambda_{12}\frac{|A_0|^2}{V}\left(
2\sum\limits_{i\ne j} n_in_j +
\sum\limits_{i} n_i(n_i-1)\right)\ .
  \end{equation}
The mean frequency shift is then given by the ratio of (\ref{H-mean-result})
and the norm (\ref{norm}):
  \begin{equation}\label{mean-shift}
\hbar\Delta\omega_{\rm coll}=\frac{\lambda_{12}}{VN}\left(
2N^2-
\sum\limits_{i} n_i(n_i+1)\right)\ .
  \end{equation}
The formal reason for the factor 2 to appear in 
Eqs. (\ref{H-mean-result}, \ref{mean-shift}) and, eventually for $g_2$ to 
appear
in Eq.(\ref{boltsshift}), is the following. In taking the average in
Eq.(\ref{H-mean}) by Wick's theorem \cite{hua63}, there are 
two essentially different ways to pair the operators, analogous
to the Hartree and Fock contributions to the energy. 
For short range interaction between bosons, 
the Hartree and Fock contributions are equal and as a result
the  frequency shift is twice as large as 
the ``mean density'' result. 

In the thermodynamic limit, $V,N\to\infty$, $n=N/V$ constant, the second term
in Eq.(\ref{mean-shift}) contributes only when there are states 
filled by a macroscopic number of particles. 
For example, in thermodynamic equilibrium
at $T<T_{\rm BEC}$, the shift Eq.(\ref{mean-shift}) 
is $\lambda_{12}(2n-n_c^2/n)$, whereas
in a non degenerate gas, at $T>T_{\rm BEC}$, the shift is $2\lambda_{12}n$. 

\noindent{\bf The sum rule:} 
We turn now to deriving a sum rule that generalizes the result 
Eq.(\ref{boltsshift})
to non-homogeneous samples and spatially varying excitation 
field (and $\lambda_{11}\ne 0$).
We start with the Golden Rule formula for the absorption
spectrum,
  \begin{equation}\label{GR}
{\cal I}(\omega) =
\frac{2\pi}{\hbar}\sum_{E_i,E_f} \delta(\hbar\omega+E_i-E_f)
|\langle f|{\cal H}_{\rm rad}|i\rangle|^2 p_i \ ,
  \end{equation}
where $|i\rangle$, $|f\rangle$ are eigenstates of the
Hamiltonian ${\cal H}={\cal H}_0+{\cal H}_{\rm int}$ with the energies $E_i$,
$E_f$, and $p_i$ is the statistical occupation of the states $|i\rangle$.

The sum rule for the spectrum ${\cal I}(\omega)$ is
found by evaluating the first moment:
  \begin{eqnarray}\label{commutator}
\int \omega {\cal I}(\omega)\frac{d\omega}{2\pi} &=&
\frac{1}{\hbar^3}\sum_{E_i,E_f} (E_f-E_i)
|\langle f|{\cal H}_{\rm rad}|i\rangle|^2 p_i \cr
&=&\frac{1}{\hbar^3}\sum_{E_i,E_f}
\langle i|{\cal H}_{\rm rad}|f\rangle
\langle f|[{\cal H}, {\cal H}_{\rm rad}] |i\rangle p_i \cr
&=& \frac{1}{\hbar^3}
\sum_{E_i}
\langle i|{\cal H}_{\rm rad} [{\cal H}, {\cal H}_{\rm rad}] |i\rangle p_i .
  \end{eqnarray}
In obtaining this result we first integrated the delta function, then wrote
the result as a matrix element of the commutator $[{\cal
H},{\cal H}_{\rm rad}]$ and, finally, used the completeness
relation.

Now we consider contributions of the  different terms of the Hamiltonian 
to the sum rule. The potential energy operator $\int
(\psi^+_1\psi_1+ \psi^+_2\psi_2)\ U(r)d^3r$ commutes with ${\cal
H}_{\rm rad}$, and thus does not contribute. 
There are two contributions, first from the interaction Hamiltonian, 
second from the kinetic energy operator, 
denoted by $F_{\rm int}$ and $F_{\rm kin}$ respectively. The sum rule becomes
 \begin{equation}\label{sum-total}
\int \omega {\cal I}(\omega)\frac{d\omega}{2\pi}=
F_{\rm int}+ F_{\rm kin} \ .
  \end{equation}
For Doppler-free excitation $F_{\rm kin}$ is  small compared to 
$F_{\rm int}$, whereas for  Doppler-sensitive excitation it contributes the 
larger shift. 

First, consider the interaction ${\cal H}_{\rm int}$, and calculate $F_{\rm 
int}$.
After evaluating the commutator in
Eq.(\ref{commutator}), one follows the same procedure as in the above 
calculation
of the norm $\parallel\Phi_1\parallel$.  The result is 
  \begin{equation}
F_{\rm{int}} =
\left\langle
\int
\left(\lambda_{12}-\lambda_{11}\right)
|A(r)|^2
\psi^+_1(r)\psi^+_1(r)\psi_1(r)\psi_1(r)\frac{d^3r}{\hbar^3}
\right\rangle ,
  \end{equation}
where $\langle ...\rangle$ means
$\sum_{E_i} \langle i| ... |i\rangle p_i $.
The expectation value 
$\left\langle :(\psi_1^+(r)\psi_1(r))^2 :\right\rangle=G_2(r)$, 
the two--particle density. (Here $:...:$
indicates canonical normal ordering.) 
Finally, using the statistical factor $g_2=G_2/n^2$, 
the result is 
  \begin{equation}\label{int}
F_{\rm int}=
\int\left(\lambda_{12}-\lambda_{11}\right)
|A(r)|^2 g_2 n^2(r)\frac{d^3r}{\hbar^3}\ .
  \end{equation}

Next, we calculate $F_{\rm kin}$, the contribution to the sum rule 
coming from the kinetic energy operator $-(\hbar^2/2m) \int
(\psi^+_1\nabla^2\psi_1 + \psi^+_2\nabla^2\psi_2) d^3r$. After
evaluating the commutator with ${\cal H}_{\rm rad}$, one has
  \begin{equation}\label{kin}
F_{\rm kin}=
-\frac{\hbar^2}{2 m} \left\langle \int \psi^+_1(r) A^*(r)
\left[\nabla^2,A(r)\right]
\psi_1(r) \frac{d^3r}{\hbar^3} \right\rangle \ .
  \end{equation}
Integrating 
by parts, and writing the excitation field as $A(r)=|A(r)|e^{i\theta}$,
yields
  \begin{equation}\label{kin-sum}
F_{\rm kin}=
\int\left( \frac{\hbar^2}{2m}|\nabla A|^2 n
-\hbar |A|^2 \vec j\cdot \vec\nabla\theta \right)
\frac{d^3r}{\hbar^3} \ ,
  \end{equation}
where $n$ and $\vec j$ are the particle number and flux densities:
  \begin{equation}\label{n-j}
n(r)=\langle\psi_1^+(r)\psi_1(r)\rangle
 , \
\vec j(r)=-\frac{i\hbar}{2m}
\langle\psi_1^+(r)\vec\nabla
\psi_1(r)\rangle + {\rm h.c.}
  \nonumber
  \end{equation}
The first term in Eq.(\ref{kin-sum}) generalizes  
the ordinary momentum recoil energy shift to the trapped gas problem\cite{cesar99}. 
The second term represents the Doppler shift due to possible macroscopic gas 
flow in the sample. To clarify this, consider 
$A(r)=A_0e^{ipr/\hbar}$, which would
describe Doppler-sensitive excitation. Then 
$F_{\rm kin}=|A_0|^2
\int(p^2/2m-\vec p\cdot\vec v)  n\ 
d^3r/\hbar^3$, where $\vec v=\vec j/n$ is the local velocity. 
The sensitivity of the frequency shift to motion
within the sample, manifest in the second term in Eq.(\ref{kin-sum}), 
makes it possible, in principle, to detect vortices
in the condensed state.  

To employ the sum rule, one needs to relate the integrated spectral power to
$A(r)$ and $n(r)$. 
Repeating the steps that led to
Eq.(\ref{sum-total}), one obtains
  \begin{equation}\label{I_tot}
{\cal I}_{\rm tot}=\int {\cal I}(\omega)\frac{d\omega}{2\pi}=
\int |A(r)|^2 n(r) \frac{d^3r}{\hbar^3}
  \end{equation}
Combining Eq.(\ref{I_tot}) with the sum rule Eq.(\ref{sum-total}), one
obtains an exact expression for the spectrum's ``center of mass'' 
$\bar\omega=\int\omega{\cal I}(\omega)d\omega/\int{\cal
I}(\omega)d\omega$.

For example, consider a uniform density sample, and ignore the spatial variation of
the laser field $A(r)$. Eq.~(\ref{int}) gives $F_{\rm int}=
\left(\lambda_{12}-\lambda_{11}\right)g_2 n^2
\int |A(r)|^2 d^3r/\hbar^3$.  In our experiment $F_{\rm{kin}}$ can be 
neglected.
Simplifying Eq.(\ref{I_tot}) and combining it with Eq.(\ref{int}) yields
the frequency shift Eq.(\ref{boltsshift}). 

There are two comments concerning the generality of the sum rule. 
First, note that in deriving the sum rule Eq.(\ref{sum-total}),
we do not assume thermodynamic equilibrium.  The result is exact and applies to
non-equilibrum systems for which the factor $g_2$ may differ from its
equilibrium value. 
Second, the above derivation of the sum rule assumes coherence of the excitation
described by (\ref{H_ext}). One can see, however, 
that the results (\ref{int}),(\ref{kin-sum}),(\ref{I_tot}) hold as well 
for an incoherent excitation field of the form 
$A(r)e^{i\omega t+i\phi(t)}$ with a fluctuating phase $\phi(t)$. 
Also, it is straightforward to generalize 
the results for the excitation field with different spatial dependence
of different frequency components.

\noindent{\bf Analysis of the data:} 
To investigate the utility of the sum rule, we applied it to
extract a value for
$a_{1S-2S}$ from data on the $1S-2S$ transition in hydrogen for a
normal gas.  An account of the
experimental situation and examples of the spectral data will be
published elsewhere \cite{ldm99}.  Evaluating the integrals in the sum
rule requires knowing
the excitation field $A(r)$, the value of $g_2$,
and the density $n(r)$.
  The excitation field is generated by a Gaussian beam which
is fully characterized by a single parameter, the beam radius, which can
be accurately determined.  For temperatures above
$T_{\rm BEC}$, $g_2 = 2$.  At lower
temperatures, $g_2$ depends on the temperature.

The density $n(r)$ was found by measuring the peak density and knowing
the properties of the trap.
The peak density $n_0$ was determined by
exploiting the property that the BEC critical density is accurately
described by the ideal gas expression:
$n_{\rm c} = 2.612 (2 \pi \hbar^2/k_B T m)^{3/2}$.
If the system is at the transition point then
measuring the temperature determines the density.
The system was cooled into the condensate regime and
the spectrum was observed as the condensate decayed.  The spectrum was
measured after the condensate had decayed for ten
seconds, when the presence of a small though visible condensate assured
that the peak density $n_0$ had its critical value. The contribution 
of the condensate to the spectrum was unimportant. Thus, the only
quantity required to apply the sum rule was the temperature.  This was
found by measuring the width of the Doppler-sensitive spectrum\cite{fried99}.

From the experimentally measured spectrum we found
$\bar\nu = -29 \pm (2)\,{\rm KHz}$.
We numerically calculated the integrals on the right hand side of
Eqs.(\ref{int}),(\ref{I_tot}) and found that
$2(\lambda_{12}-\lambda_{11})/h
= 4.4 \pm(1.7)\times 10^{-10}\,{\rm Hz\, cm^3}$, where the major sources of
uncertainty are the temperature and the trap and laser geometry.
From this we determined the  $1S-2S$ scattering
length to be $a_{12} = -1.6\pm(0.7)\,{\rm nm}$.  This result is in reasonable
agreement with the calculated value \cite{jdd96}, $a_{12}= - 2.3\,{\rm nm}$.

An alternative approach to extracting the scattering length
was used in Ref.~\cite{kfw98}, where the value $a_{12}= - 1.4 \pm
(0.3)\,{\rm nm}$ was reported.  The interpretation
employed
a semiclassical description of the atomic motion and a local density
description of the phase shift.  The present method is
more direct and, we believe, more reliable.  It can be viewed as a
check on the earlier analysis, and a confirmation of the calculation
of the dipolar decay constant\cite{dipolar} on which it depends.

In reference \cite{fkw98}, an internally consistent description of the
density of the condensate required assuming that the density shift
parameter in the condensate was the same as in the normal gas,
rather than half as large as
expected from this analysis. This anomaly remains to be explained.

As a speculative explanation, one could consider 
a state close to the transition temperature in dynamical but not in thermal
equilibrium, in which several low energy states are populated with macroscopic 
occupation numbers. For example,
for $N$ particles distributed equally among $m$ states, 
one has $n_i=N/m$, $i=1,...,m$ in Eq.(\ref{H-mean-result}). 
Then from Eq.(\ref{mean-shift}) $\hbar\Delta\omega_{\rm 
coll}=(2-1/m)\lambda_{12}n$,
{\it i.e.},
the shift is described by an effective $g_2=2-1/m$. 
For a large number $m$ of constituent states, 
the effective $g_2$ can be arbitrarily close to $2$.


In summary,
we have shown that quantum statistical
correlations of a cold gas sample are imprinted in the collisional shift of
the center of mass of an optical absorption spectrum.
In the cold collision regime
the sum rule Eq.(\ref{sum-total})
can be applied to determine the statistical
correlation factor $g_2$
from optical spectrum.
The sum rule is valid for any gas in
the cold collision regime.  It takes into account possible
inhomogeneities in the
sample and the excitation field, and it is valid above and below $T_{\rm  
BEC}$. Also, the sum
rule is valid for a non-equilibrium system, with $g_2$
values possibly different from those in equilibrium. We have demonstrated
the usefulness of the sum rule by using it to extract the
$1S-2S$ scattering length for hydrogen from experimental data.

It should be pointed out that our results, the frequency 
shift~(\ref{boltsshift}) and the sum rule~(\ref{sum-total}),
are only valid at small mixing angles of the $1S$ and $2S$ states. 
The cold collision shift at large angle mixing is an important problem, 
particularly for atomic clocks. The generalization of the results (\ref{sum-total})
and (\ref{boltsshift}) for such systems is an interesting open problem.

\acknowledgements
\noindent
We thank D.~Fried, L.~Willman, D.~Landhuis and S.~Moss 
for their contributions in obtaining and analyzing the data.
We thank T.~J.~Greytak and W.~Ketterle for helpful conversations.  
The experimental work was supported by the National Science Foundation 
and the Office of Naval Research.

\end{document}